\documentclass[prd,nofootinbib]{revtex4}

\usepackage{graphicx}
\usepackage{amssymb}
\usepackage{epstopdf}
\usepackage{subfigure}

\def\ba{\begin{array}}
\def\ea{\end{array}}
\def\be{\begin{equation}}
\def\ee{\end{equation}}
\def\ben{\begin{equation} \nonumber}
\def\een{\end{equation}}
\def\baray{\begin{eqnarray*}}
\def\earay{\end{eqnarray*}}

\begin{document}
\title{Cosmic Strings as the Source of Small-Scale Microwave Background Anisotropy}
\author{Levon Pogosian$^{1,2}$, S.-H. Henry Tye$^3$, Ira Wasserman$^4$ and Mark Wyman$^2$}
\affiliation{$^1$Department of Physics, Simon Fraser University, 8888 University Drive, Burnaby, BC, V5A 1S6, Canada. \\
$^2$Perimeter Institute for Theoretical Physics, \\ 31 Caroline St. N, Waterloo, ON, L6H 2A4, Canada. \\
$^3$Laboratory for Elementary Particle Physics,
Cornell University, Ithaca, NY 14853, USA. \\
$^4$Center for Radiophysics and Space Research, Cornell
University, Ithaca, NY 14853, USA}


\date{today}

\begin{abstract}
Cosmic string networks generate cosmological perturbations actively throughout the history of the universe. Thus, the string sourced anisotropy of the cosmic microwave background is not affected by Silk damping as much as the anisotropy seeded by inflation. 
The spectrum of perturbations generated by strings does
 not match the observed CMB spectrum on large angular scales ($\ell<1000$) and is bounded to contribute no more than $10\%$ 
of the total power on those scales. However, when this bound is marginally saturated, the anisotropy created by cosmic strings on small angular scales
$\ell \gtrsim 2000$ will dominate over that created by the primary inflationary
perturbations. This range of angular scales in the CMB is presently being measured by a number of experiments; their results will test this prediction of cosmic string networks soon.
\end{abstract}
\maketitle

\section{Introduction}

In the past decade, a series of experiments, in particular the
Wilkinson Microwave Anisotropy Probe (WMAP) \cite{Dunkley:2008ie}, 
have measured the anisotropy in the 
cosmic microwave background (CMB) radiation with extraordinary precision. The data from these experiments have helped to usher in the era of precision cosmology. Most of the cosmological results that have been obtained from these experiments have been derived from the microwave anisotropy at relatively large angular scales. The remarkable agreement between the angular power spectrum of these data and the predictions of the adiabatic inflationary scenario have established the empirical success of inflationary cosmology. 

Cosmic strings, though ruled out as the origin of cosmological structure,
have recently enjoyed a renaissance. This renewed popularity 
has been brought about by the recognition that a variety of string theory-motivated
and hybrid models for inflation generically
predict the formation of cosmic string networks 
\cite{Tye:2006uv,Vilenkin,kofman,tkachev,Rjeannerot,mairiS,cop, Jeannerot:2006dy,jst,costring,DV03,Polchinski03,mcgill,Firouzjahi:2005dh}.
Strings are limited to producing less than about 10\% of the
primordial CMB anisotropy \cite{contaldi,battye,bouchet,bevis,Wyman:2005tu,Seljak:2006bg,Bevis:2007gh}, though it was shown in \cite{Bevis:2007gh,Battye:2007si} that CMB data can actually favor a contribution from strings if the inflationary spectrum is exactly Harrison-Zeldovich ($n_s=1$);
this corresponds to a string tension ($G\mu$) between $4 \times10^{-7}$ and $6\times10^{-7}$
for a standard set of string network parameters. 

The B (i.e., curl) mode polarization in the CMB caused by the 
active perturbations of a cosmic string network 
has a spectrum distinct from those expected either from inflationary 
gravity waves or the lensing of E (i.e., gradient) mode polarization into B mode by 
large scale structure. When the above bound is marginally saturated, this B mode polarization
should be measurable \cite{Turok:1997gj,Pogosian:2003mz,Seljak:2006hi,Bevis:2007qz}, providing a powerful test of the presence of cosmic strings. 
Another consequence is the power spectrum of perturbations
that strings source at large $\ell$ (small angular size).
Cosmic string networks continually generate CMB anisotropies, both primordially through active
density perturbations and subsequent to recombination through the lensing of the primary
CMB light -- the Kaiser-Stebbins effect (KS) \cite{Kaiser:1984iv}.
The CMB anisotropy due to the KS effect alone at large $\ell$ is expected to decrease only as $\sim1/\ell$ \cite{Hindmarsh:1993pu}.
This rate of decrease is much slower than that expected for 
inflationary perturbations (which fall off exponentially as a function of $\ell$ 
due to Silk damping, which is due to radiative diffusion).
If $G \mu$ is not too small, this large $\ell$ power spectrum may be measurable. It should appear as an excess above the prediction from inflationary perturbations.
In particular, for $G\mu \approx 3\times 10^{-7}$, the power created on small angular scales, $\ell \gtrsim 2000$, by cosmic strings will actually dominate over that created by the primary inflationary
perturbations. This range of angular scales in the CMB is presently being measured by a number of experiments, so that this prediction of cosmic string networks will be tested soon. In this note, we present the large $\ell$ power spectrum due to cosmic strings.

\section{Cosmic String Model}

We use CMBACT \cite{cmbact,Pogosian:1999np}, a modified version of CMBFAST \cite{cmbfast}, to produce the string sourced anisotropy spectra. The model, described in Refs.~\cite{Pogosian:1999np,Wyman:2005tu,our_erratum}, is based on representing the cosmic string network
as a collection of moving straight string segments. In brief, there are two
important length scales in this model: $\xi$, the length 
of a string segment, which represents the typical length
of roughly straight segments in a full network; and
$\bar{\xi}$, the typical length between two string segments,
which sets the number density of strings in a given volume ($N_s\propto1/\bar{\xi}^2$).
We use the Velocity-dependent
One-Scale Model (VOS) \cite{MS96} to evolve the network length-scales and density. In the simplest and most relevant cases, $\xi = \bar{\xi}$.
With these parameters set, we average over a set of randomized
realizations of the network approximated as a set of straight string segments.
This model was introduced in \cite{ABR97}, based on the approach suggested in \cite{Vincent:1996qr}, and was developed into its present form in \cite{Pogosian:1999np,Gangui:2001fr}.
The overall normalization of the spectrum has a simple dependence
on the string tension and number density:
\be
C_\ell^{\rm strings} \propto N_s (G\mu)^2 \propto \left ( \frac{G\mu}{\bar{\xi}} \right)^2.
\ee
Lower reconnection probability \cite{Jackson:2004zg}
for cosmic strings will rescale the amplitude of this spectrum, but will not change its shape. 
Although this model assumes a single tension cosmic string network, the results described here should generally apply for more complex multi-tension string network models of the sort that may be produced in the aftermath of brane inflation \cite{Polchinski03, Tye:2005fn, Leblond:2007tf}. 

\begin{figure}[tbp] 
{\includegraphics[width=3in]{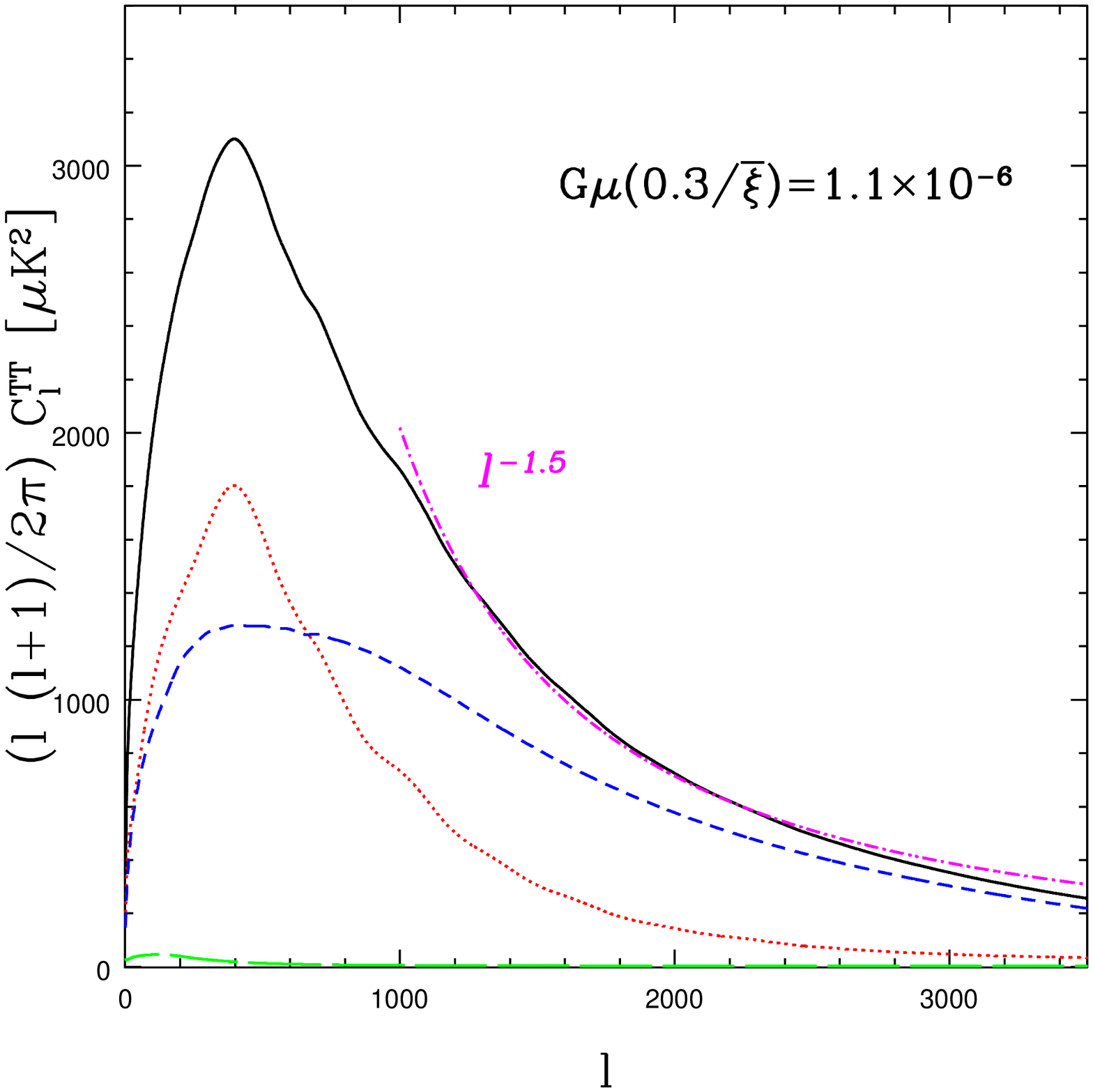}}
{\includegraphics[width=3in]{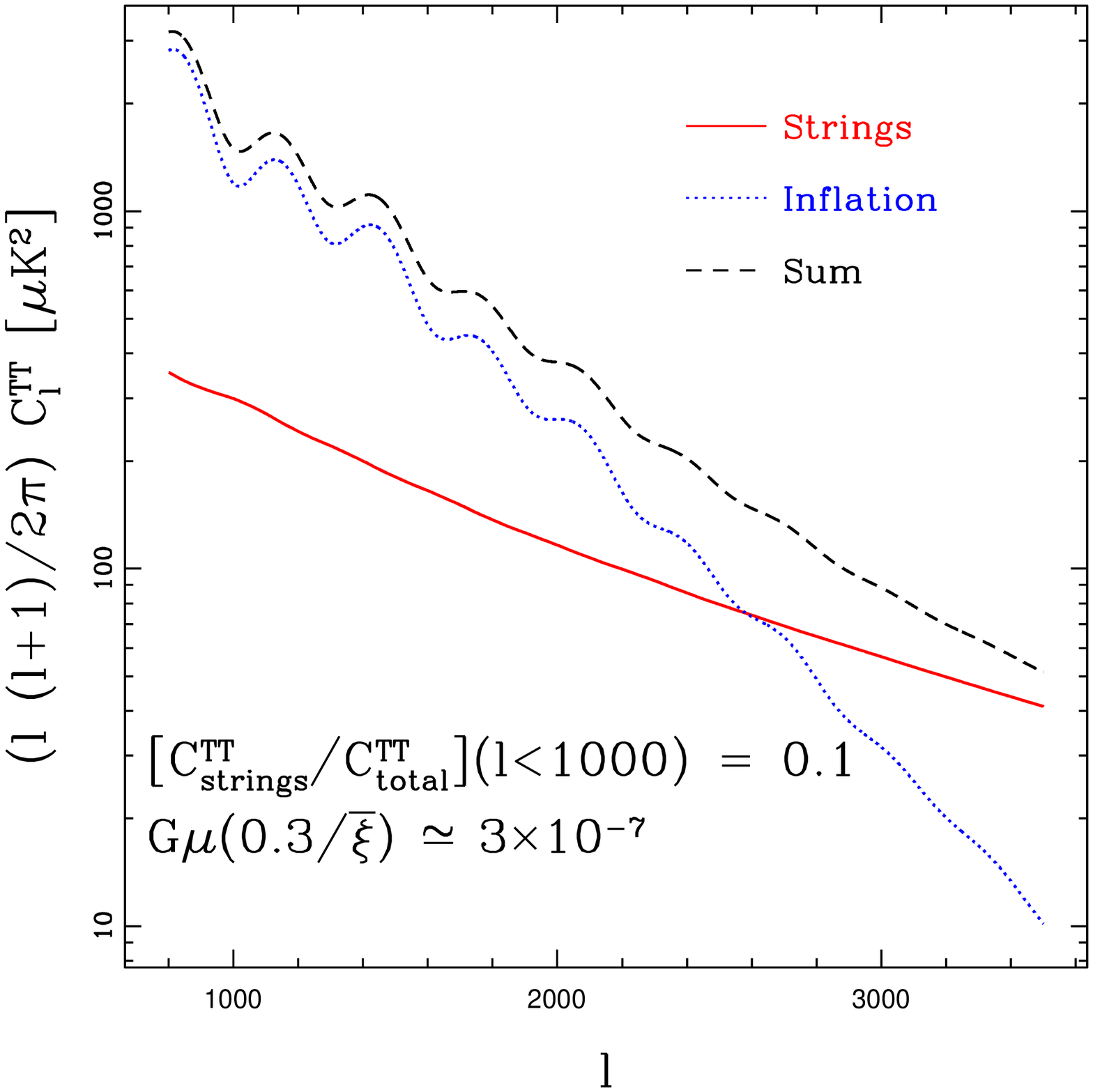}}
\caption{ {\it Left:} The TT power sourced by a cosmic string network with
 $G\mu =1.1\times10^{-6}$; 
average coherence length $\xi \simeq 0.2$ (measured in units of the horizon size); rms 
velocity near $v=0.2c$; and average wiggliness parameter $\alpha=1.3$. The solid black line is the total power. The red dotted line is the power due to scalar perturbations, the blue dashed line represents the vector mode perturbations. The (green) long dashed line shows the (negligible) contribution from tensor modes. Note that vector modes dominate above $\ell \simeq 800$, a distinctively stringy effect.
The magenta dash-dotted line shows the $\ell^{-1.5}$ fitting formula.
{\it Right:} Contributions from strings (solid red), inflation (blue dot) (including lensing, from CAMB \cite{Challinor:2005jy}), and their combination (black dash) at high-$\ell$. The string contribution satisfies the $10$\% bound on total power imposed on large angular scales.}
\label{f1}
\end{figure}

Fig.~\ref{f1} shows our result. The dominance of vector modes (the blue dashed line in the left panel) for $\ell \gtrsim 800$ is an effect that is a distinctive signature of cosmic strings. As we show with the magenta, dash-dotted line, the string spectrum at $1000< \ell < 3500$ -- where 
vector-sourced perturbations start to dominate -- is well approximated
by a $\ell^{-1.5}$ fall off. This can be captured by a simple fitting formula:
\be
(\ell (\ell+1)/ 2\pi) C_\ell^{TT}[\rm{strings}] \simeq 500 \left ( \frac{2000}{\ell} \right )^{1.5}  \left ( \frac{G\mu}{1.1\times10^{-6}} \right )^2 \quad \quad \rm{for } \; \ell \gtrsim 1000.
\ee
We find that towards the very high $\ell$ end of the considered range, i.e. at $\ell > 3000$, the fall off is better described by $\ell^{-1}$, in agreement with the pure KS contribution analytically predicted in \cite{Hindmarsh:1993pu}. At smaller $\ell$, the residual fluctuations from the last-scattering surface are non-negligible leading to a $\ell^{-2}$ fall off in the $1000<\ell<2000$ range.

The right panel in Fig.~\ref{f1} shows the high-$\ell$ power sourced by strings relative to the inflationary contribution.
The string spectrum's amplitude is set by saturating the observational bound:
it accounts for $10$\% of power for $\ell<1000$.

The spectrum in Fig.~\ref{f1} should not be taken as the unique prediction of the $C_\ell$ spectrum
from strings, but as a representative example of what one can get for a reasonably 
motivated string network.  It corresponds to particular values of 
string model parameters, such as wiggliness, coherence length, and rms 
velocity. It also relies on the ``moving segments" approximation used in
CMBACT. This model is designed to describe
statistical properties of scaling string networks. For instance, it can
be used to calculate CMB power spectra, but cannot make a sky map
of string network effects.  Real string
simulations that capture more of the physics of networks-- such as their curvature and
loops -- are too computationally costly to be used over many
expansion times.  The segments model has been shown to match CMB spectra
from full simulations reasonably
well over the scales where they can be compared (for a fuller discussion, see Refs.~\cite{Pogosian:1999np,Wyman:2005tu,our_erratum}). 

While varying the string tension, $G\mu$, simply renormalizes 
the spectrum, changing the other parameters can re-distribute the 
power between the large and small scales. Generally, smaller velocities 
and smaller coherence lengths enhance the power in vector modes at high 
$\ell$ \cite{Pogosian:2007gi}.  More wiggliness tends to make 
strings move more slowly. This suppresses all contributions to the anisotropic 
stress, including the vector modes \cite{Pogosian:1999np,Pogosian:2007gi}. 
For this plot, we used a model with $G\mu =1.1\times 
10^{-6}$ with an average string coherence length of $\sim 0.2$ of the horizon size \footnote{The coherence length of infinite strings is typically of the order of the horizon size at each epoch.
A smaller {\it average} coherence scale comes about when a significant fraction of the network density is present in the form of large string loops. The fraction of string density in loops grows for smaller values of $G\mu$.}, a rms velocity of $\sim0.2c$, and an average wiggliness parameter $\alpha$ of $1.3$. 
These values are fairly close to those seen in numerical simulations (e.g. \cite{Allen:1990tv}),
 but are slightly different from 
those used in the default version of CMBACT; the values chosen here are those which slightly
enhance the string-sourced power at high $\ell$. Fractionally greater (or attenuated) power at small
angular scales can be achieved with different model parameters, but the overall
behavior at high $\ell$ is a generic feature of string networks.
All cosmological parameters, \emph{i.e.} $\Omega_M$, $h$, etc, are those of the latest WMAP best fit \cite{Dunkley:2008ie}.

\section{Conclusion and Remarks}
\begin{figure}[tbp] 
   \includegraphics[width=3in]{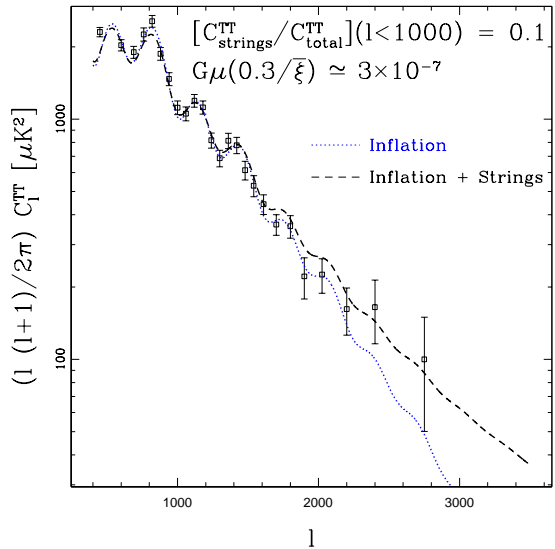} 
   \caption{The predicted high $\ell$ CMB anisotropy in the ACBAR range, plotted
   against the latest data from the ACBAR experiment. The dotted 
   blue line is the inflationary prediction alone (including lensing, from the software
   package CAMB \cite{Challinor:2005jy}); the dashed black line includes the
   contribution from a cosmic string network. The amplitude fitting has been 
   done roughly, by eye, to match the two theoretical curves
   with the second ACBAR data point; the cosmological parameters are all those of the WMAP best fit. }
   \label{withacbar}
\end{figure}

Cosmic strings produce power on small angular scales because they
are \emph{active} sources that continue contributing to the anisotropy after the last scattering. For that reason they evade Silk damping, \emph{i.e.} the erasure of anisotropies on small scales due to the finite thickness of the last scattering surface.
A network of cosmic strings with a tension near the present observational
bound of $G \mu \lesssim 6 \times 10^{-7}$ can dominate the power
spectrum of CMB fluctuations
in the strongly Silk-damped regime ($\ell > 2000$) of the microwave background
anisotropy, creating an apparent excess of power over what is 
expected from an inflationary adiabatic perturbation spectrum. 
This high $\ell$ regime is accessible to existing fine scale resolution  
experiments like the Cosmic Background Imager (CBI) \cite{readhead} and
the Arcminute Cosmology Bolometer Array Receiver (ACBAR) \cite{ACBAR} now,
and will soon be measured very accurately by experiments like the South Pole Telescope \cite{SPT}
and the Atacama Cosmology Telescope \cite{ACT}. 
It is interesting to note that these experiments' published data already show
some hints of excess power in the high-multipole range; see Fig. \ref{withacbar}
for a rough comparison with the data from ACBAR.
Although we have not done a statistical analysis, the preliminary
results presented here suggest that cosmic strings with $G\mu(0.3/\xi)\simeq
3\times 10^{-7}$ could contribute enough power to account for the excess
over inflation suggested by the ACBAR data at $\ell\gtrsim 2000$ without
exceeding observational bounds on string contributions at $\ell\lesssim 1000$.
Excess power in the high-$\ell$ CMB can also be generated by other
physical phenomena,
including the Sunyaev-Zeldovich effect \cite{szeffect}
and tangled primordial magnetic fields \cite{magnetic}.
Using the SZ effect to account
for any substantial excess measurable by today's experiments is problematic, however, 
because the amount of small-scale gravitational clustering (measured by
the parameter $\sigma_8$) required
to generate a large excess over the inflationary prediction via the SZ effect
is in some conflict with values determined by other experiments.
Tangled magnetic fields, on the other hand, are not meaningfully constrained by  
competing experiments,
but their existence at the necessary epoch is by no means accounted for.
Discovery of significant high $\ell$ excess power in CMB that cannot  
be explained by more conventional means could be taken as evidence 
for existence of  cosmic
strings with tensions near the observational bound. 
On the other hand, if no excess is seen at large $\ell$,  
non-observation of this effect will provide a useful bound 
on the properties of any cosmic string network.
Strings with tensions in this range can be  
searched for by some other means, such as gravitational lensing and 
microlensing \cite{lensing},  
gravitational radiation bursts \cite{gravitywaves}, pulsar timing or non-Gaussian step-like fluctuations  
in CMB temperature \cite{steps}.
An especially promising signature of such strings would be a substantial  
B mode polarization \cite{Turok:1997gj,Pogosian:2003mz,Seljak:2006hi,Bevis:2007qz}; 
for $G\mu$ around $6\times 10^{-7}$, the
B mode polarization fluctuations from strings could exceed expected power from E to B conversion by  
gravitational lensing by factors of a few.

\acknowledgments 
We thank J. Richard Bond, Anthony Readhead, and Mark Hindmarsh for discussions leading to the 
exploration of this question; we also thank David Chernoff for conversations and
Jonathan Sievers for comments on the draft.
LP's work is supported by a Discovery grant from the National Sciences and Engineering Research Council of Canada. SHHT's work is supported by the National Science Foundation (NSF) under grant PHY-0355005. IW's work is supported by the NSF under grant PHY-0555216. The work of M.~W. at the Perimeter Institute is supported by the Government of Canada through Industry Canada and by the Province of Ontario through the Ministry of Research \& Innovation.

\end{document}